\documentclass[11pt]{article}
\pdfoutput=1
\usepackage{graphicx}
\usepackage[margin=1in]{geometry}
\usepackage{caption}
\usepackage{subcaption}
\usepackage{amsmath}
\usepackage{amssymb}
\usepackage{MnSymbol}

\usepackage{color}

\newcommand{\fNL}{f_{\rm NL}}
\newcommand{\bk}{\mathbf{k}}
\newcommand{\bx}{\mathbf{x}}
\newcommand{\skewR}{\mathcal{M}_{3,R}}
\newcommand{\fNLloc}{f_{\rm NL}^{\rm local}}
\newcommand{\fNLeq}{f_{\rm NL}^{\rm equil}}
\newcommand{\fNLorth}{f_{\rm NL}^{\rm orthog}}

\begin{document}

\begin{titlepage}
\begin{center}

{ \huge \bfseries Influence of large local and non-local bispectra on primordial black hole abundance}\\[1cm]

Sam Young$^{1}$, Donough Regan$^{2}$, Christian T. Byrnes$^{3}$\\[0.5cm]
Department of Physics and Astronomy, Pevensey II Building, University of Sussex, BN1 9RH, UK\\[0.5cm]
$^{1}$S.M.Young@sussex.ac.uk, $^{2}$D.Regan@sussex.ac.uk, $^{3}$C.Byrnes@sussex.ac.uk \\[1cm]

\today\\[1cm]

\end{center}

Primordial black holes represent a unique probe to constrain the early universe on small scales - providing the only constraints on the primordial power spectrum on the majority of scales. However, these constraints are strongly dependent on even small amounts of non-Gaussianity, which is unconstrained on scales significantly smaller than those visible in the CMB. This paper goes beyond previous considerations to consider the effects of a bispectrum of the equilateral, orthogonal and local shapes with arbitrary magnitude upon the abundance of primordial black holes. Non-Gaussian density maps of the early universe are generated from a given bispectrum and used to place constraints on the small scale power spectrum. When small, we show that the skewness provides an accurate estimate for how the constraint depends on non-Gaussianity, independently of the shape of the bispectrum. We show that the orthogonal template of non-Gaussianity has an order of magnitude weaker effect on the constraints than the local and equilateral templates.

\end{titlepage}


\section{Introduction}

Primordial black holes (PBHs) are black holes which may have formed very early on in the history of the universe from the collapse of density perturbations generated during inflation. During inflation, quantum fluctuations are stretched out by the rapid expansion of the universe, and quickly become larger than the Hubble horizon, becoming classical density perturbations. Once inflation ends, the perturbations begin to reenter the horizon, and if large enough, can collapse to form a PBH. Because such perturbations can reenter the horizon before baryogenesis, there is no need for such black holes to have a large enough mass to overcome degeneracy pressures - and the formation of PBHs with very small masses is possible.

Because PBHs form on small scales, they have often been used to constrain the smallest scales in the early universe. Precision measurements and constraints upon the primordial Universe are available from the cosmic microwave background (CMB) and large scale structure (for example, the constraints on inflation from Planck \cite{Ade:2013uln}), but these only provide constraints on the largest 6-8 e-folds inside the visible universe - while inflation is expected to have lasted 50-60 e-folds. PBHs, on the other hand, provide constraints on a much greater range of scales, spanning around 50 e-folds, although the constraints are much weaker.

Many attempts have been made to detect PBHs, yet they remain undetected. However, a tight upper limit can be placed on the abundance of PBHs, which is typically stated in terms of the mass fraction of the Universe contained within PBHs at the time of formation, $\beta$. Constraints on $\beta$ vary greatly for PBHs of different mass, ranging from $\beta<10^{-5}$ to $\beta<10^{-25}$. For a summary of the constraints see \cite{Carr:2009jm}. Because the number of PBHs forming depends on the primordial power spectrum, constraints on the abundance of PBHs can be used to place bounds on the power spectrum \cite{Josan:2009qn}. These constraints on the power spectrum are typically of order $10^{-2}$, significantly weaker than constraints from the CMB.

In order for a significant number of PBHs to form, the power spectrum needs to be orders of magnitude larger than is observed in the CMB - meaning that it must become large on small scales. There are a range of models for inflation which do predict such behaviour, whilst being consistent with current cosmological observations. Such models include the running mass model \cite{Kohri:2007qn,Drees:2011hb}, axion inflation \cite{Bugaev:2013fya}, a waterfall transition during hybrid inflation \cite{Bugaev:2011wy, Lyth:2012yp,Halpern:2014mca}, from particle production during inflation \cite{Erfani:2015rqv}, inflationary models with small field excursions but which are tuned to produce a large tensor-to-scalar ratio on large scales \cite{Hotchkiss:2011gz}, and can be formed from passive density fluctuations \cite{Lin:2012gs}. See also \cite{Linde:2012bt,Torres-Lomas:2014bua,Suyama:2014vga}. For further reading and a summary of various models which can produce PBHs, see \cite{Green:2014faa}. Such models typically predict at least a small amplitude of non-Gaussianity - and it has been shown that constraints on the small scale power spectrum are strongly dependent on non-Gaussianity \cite{Byrnes:2012yx} - and can vary by over an order of magnitude. 

Previous papers have used an analytic method to investigate the effects of non-Gaussianity - and were limited to either investigating local-type non-Gaussianity for which analytical results are available \cite{Bullock:1996at,Ivanov:1997ia,PinaAvelino:2005rm,Seery:2006wk,Shandera:2012ke,Lyth:2012yp,Byrnes:2012yx,Young:2013oia} or, in the case of Shandera et al \cite{Shandera:2012ke}, also to a small amplitude of equilateral non-Gaussianity. This paper goes beyond previous work to investigate the effects of three different bispectrum shapes of arbitrary size on the abundance of PBHs, and on the resulting constraints, by making use of non-Gaussian density maps. We make the first study of orthogonal non-Gaussianity, and show that, for a given value of $\fNL$, it has a much smaller effect on the constraints than the equilateral and local non-Gaussian templates. We explain this observation by calculating the skewness parameter as a function of all three bisepectral templates.


The paper is organised as follows: in section \ref{sec:sim_proc} the generation of the density maps and calculation of the PBH abundance is detailed. In section \ref{sec:constraints}, the constraints on the power spectrum as a function of the bispectrum are calculated. Section \ref{sec:conclusions} concludes with a summary of the results.

\section{Simulation Procedure}\label{sec:sim_proc}
\subsection{Generation of non-Gaussian density maps}

Methods for the simulation of a map incorporating an arbitrary bispectrum were developed by Regan et al. in a series of papers \cite{Fergusson:2010ia,Regan:2011zq,Schmittfull:2012hq} (see also \cite{Wagner:2010me}). Representing the primordial curvature in Fourier space as $\zeta(\bk)$, one may simulate the curvature of a Gaussian distribution using a random number generator with variance per scale, $k$, given by the power spectrum $P_\zeta(k)$ (and zero mean). For clarity we will denote the Gaussian map as $\zeta_G(\bk)$. The bispectrum $B_\zeta(k_1,k_2,k_3)$, given by the expectation value of the three point function
\begin{equation}
\langle \zeta(\bk_1)\zeta(\bk_2)\zeta(\bk_3)\rangle =(2\pi)^3 \delta_D(\bk_1+\bk_2+\bk_3) B_\zeta(k_1,k_2,k_3)\,,
\label{eq:bispdef}\end{equation}
may be simulated using the Gaussian maps by calculating $\zeta(\bk)=\zeta_G(\bk)+\fNL \zeta_B(\bk)$  where
\begin{equation}
\zeta_B(\bk)=\int \frac{d^3\bk_1}{(2\pi)^3} \frac{d^3\bk_2}{(2\pi)^3}\delta_D(\bk-\bk_1-\bk_2)\zeta_G(\bk_1)\zeta_G(\bk_2)\frac{B_\zeta^{\fNL=1}(k,k_1,k_2)}{2\left(P_\zeta(k)P_\zeta(k_1)+P_\zeta(k)P_\zeta(k_2)+P_\zeta(k_1)P_\zeta(k_2)\right)}\,.
\end{equation}
Here we define the quantity $\fNL\equiv 5 B_\zeta(k,k,k)/(18 P_\zeta(k)^2)$ such that $B_\zeta \equiv\fNL B_\zeta^{\fNL=1}$.
Direct implementation of this convolution is numerically prohibitive unless the bispectrum can be written in a separable form, i.e.~in the form $f(k_1)g(k_2)h(k_3)$ for arbitrary one dimensional functions $f,g,h$. This is possible for sufficiently smooth generic bispectra using techniques developed in \cite{Fergusson:2009nv,Regan:2010cn,Fergusson:2010gn}. In particular, a partial wave decomposition may be employed to write the bispectrum in the form
\begin{equation}
\frac{B_\zeta^{\fNL=1}(k_1,k_2,k_3)}{2\left(P_\zeta(k_1)P_\zeta(k_2)+P_\zeta(k_1)P_\zeta(k_3)+P_\zeta(k_2)P_\zeta(k_3)\right)} = \sum_{rst} \alpha_{rst}^Q q_{\{ r}(k_1) q_s(k_2) q_{t\}}(k_3)\,,
\end{equation} 
where the notation $\{ r s t\}$ refers to all symmetrised combinations of the labels $r,s,t$ - necessary due to symmetry of the bispectrum. The triple label indices may be partially ordered such that a single index $n\equiv \{ r s t\}$ may be used to enumerate the coefficients of the expansion in the form $\alpha_n^Q$. Calculation of these coefficients only requires an inner product on the space of bispectra - restricted due to the triangle condition imposed by the Dirac delta condition in~\eqref{eq:bispdef}. Interested readers are referred to  \cite{Regan:2013wwa} for further details of the decomposition procedure. Given this decomposition, calculation of the bispectrum map reduces to calculation of fast Fourier transforms with
\begin{align}
\zeta_B(\bk)&=\frac{1}{2}\sum_n \alpha_n^Q \int d^3 \bx e^{i\bk\cdot\bx} q_{\{ r}(k) M_s(\bx) M_{t\}}(\bx)\,,\\
{\rm where} \quad M_s(\bx)&=\int \frac{d^3\bk}{(2\pi)^3}e^{i\bk\cdot\bx} q_s(k) \zeta_G(\bk)\,.
\end{align}

In this paper we focus on the three standard bispectrum templates (local, equilateral, orthogonal) for which the respective bispectra are of the form
\begin{align}
B_\zeta^{\rm local}(k_1,k_2,k_3)&= \frac{6}{5}\fNL \left(P_\zeta(k_1)P_\zeta(k_2)+ P_\zeta(k_1)P_\zeta(k_3)+P_\zeta(k_2)P_\zeta(k_3)\right)\,,\\
B_\zeta^{\rm eq}(k_1,k_2,k_3)&= \frac{18}{5}\fNL \Big(-\left[P_\zeta(k_1)P_\zeta(k_2)+ 2\, {\rm perms}\right]- 2\left[ P_\zeta(k_1)P_\zeta(k_2)P_\zeta(k_3)\right]^{2/3} \nonumber\\
&\qquad\qquad\,\, +\left[P_\zeta^{1/3}(k_1)P_\zeta^{2/3}(k_2)P_\zeta(k_3)+ 5\, {\rm perms}\right]\Big)\,,\\
B_\zeta^{\rm orth}(k_1,k_2,k_3)&= \fNL \Big(3 B_\zeta^{\rm eq,\fNL=1}(k_1,k_2,k_3) -\frac{36}{5}\left[ P_\zeta(k_1)P_\zeta(k_2)P_\zeta(k_3)\right]^{2/3} \Big) .
\end{align}
For clarity we will, where necessary, distinguish $\fNL$ for the various shapes by writing $\fNL^{\rm local}, \fNL^{\rm eq}$ and $\fNL^{\rm orth}$. We note that for the local model, the map making procedure reduces to the simple form
\begin{equation}
\zeta(\bk)=\zeta_G(\bk) + \frac{3}{5}\fNL (\zeta_G\star\zeta_G)(\bk)\, ,
\label{localtype}
\end{equation}
where the symbol $\star$ indicates a convolution.

Our simulations are carried out on a grid of $128^3$ points, and employ a scale invariant power spectrum of the form $P_\zeta(k) = A_\zeta/k^3$. The amplitude $A_\zeta$ is given by the Planck value $A_\zeta=4.75\times 10^{-8}$ but is boosted for either one efold of points or 2.5 efolds\footnote{Given our use in this paper of $128$ grid points in each dimension, we boost those values corresponding to the one efold (respectively, 2.5 efold) range $k\in [10,27.2]\Delta k$ (respectively, $k\in [10,128]\Delta k$) where $\Delta k$ is the resolution size of the grid. The lower limit, $10\Delta k$ is chosen for numerical stability to be significantly larger than the resolution size.} to a much larger amplitude - typically of order $10^{-2}$ - required to form a significant number of PBHs; the boosted region of the power spectrum will be referred to as the peak in the power spectrum later in the paper. The amplitude of this boost is then tuned such that the required amount of PBHs would form. Calculation of the PBH abundance is discussed in the following section.

We restrict our analysis to the bispectrum, but note that generating a non-zero bispectrum inevitably results in non-zero higher n-point functions. For the local model, this corresponds to generating the minimum possible trispectrum with $\tau_{NL}=(6 f_{\rm NL}/5)^2$ and $g_{\rm NL}=0$ \cite{Byrnes:2006vq}, which Shandera et al  call the hierarchical scaling \cite{Shandera:2012ke}. Our simulations automatically take this into account. However, care should be taken in interpreting the large $f_{\rm NL}$ regime for the equilateral and orthogonal models for which the trispectrum may be of a different form.

\subsection{Calculation of PBH abundance}
As described in \cite{Young:2014ana} the abundance of primordial black holes should be computed using the density contrast rather than the primordial curvature perturbation, due to the damping of super-horizon modes by a factor $k^2$. In addition, it is necessary to account for the window function $W(R,x)$ with which the density contrast is smoothed on a given scale $R$. The formation criteria is typically expressed in terms of the average over-density at the time of horizon crossing (and the PBH forms shortly after) - and so the smoothing scale $R$ corresponds to the horizon scale. As the horizon scale is the only physical scale in the Universe at this time, it is taken to be of arbitrary size, and other lengths are defined relative to the horizon/smoothing scale, $R$. The simulation does not correspond to a specific scale and the constraints on the power spectrum obtained can therefore be applied to any physical scale.

Assuming radiation domination, the relationship between the smoothed density fluctuation, $\Delta_R$, and the curvature perturbation, $\zeta$, is given by
 \begin{equation}
 \Delta_R(\bx) = \int \frac{d^3 \bk}{(2\pi)^3} e^{i\bk.\bx}\tilde{W}(R,k) \frac{4}{9} ( k R)^2 \zeta(\bk)\,,
 \end{equation}
where $\tilde{W}(R,k)$ denotes the Fourier transform of the window function. In this work we employ a volume-normalised Gaussian window function, such that\footnote{We shall drop the tilde in what follows and assume the window function is in Fourier space unless otherwise specified.}
\begin{equation}
\tilde{W}(R,k) = \exp\left(-\frac{k^2 R^2}{2}\right)\,.
\end{equation}

In order to compute the abundance, $\beta$, of PBHs we count the number of grid points for which the smoothed density exceeds the threshold, $\Delta_c$ at which PBHs form, i.e. such that $ \Delta_R(\bx) > \Delta_c$\,. Our computation of the variance is performed by Fourier transforming the smoothed density contrast to real space to obtain $\Delta_R(\bx)$, and then calculating
\begin{equation}
\mathcal{P}_{\Delta_R}=\langle \Delta_R(\bx)^2 \rangle\,,
\end{equation} 
where $\langle \dots \rangle$ represents the averaging over all grid points in real space. For ease of comparison to the literature, which do not employ a smoothing function, we note that for the rescaled density contrast, $\tilde{\Delta}_R=\exp(1/2)\Delta_R$, we obtain the approximate result $\mathcal{P}_{\tilde{\Delta}_R}\approx (4/9)^2\mathcal{P}_{\zeta}$ due to the function $(kR)^2 {W}(R,k)$ peaking with value $\exp(-1/2)$ in the boosted region. We will make use of this (accurate) approximation in the remainder of this paper. The threshold at which
PBHs form at any grid point $\bx$ is taken to be $\tilde{\Delta}_c\equiv \exp(1/2){\Delta}_c= 4/9$. This corresponds to a threshold $\Delta_c\simeq 1/3$, as used in previous theoretical predictions - though is slightly below the accepted value $0.45$ calculated from simulations \cite{Shibata:1999zs,Musco:2008hv,Nakama:2013ica}.

The variance of the Gaussian density map - denoted $\sigma$ for clarity of notation - may be evaluated as
\begin{equation}
\sigma^2 = \int \frac{dk}{k}\frac{A_\zeta(k)}{2\pi^2} (k R)^4\frac{16}{81}{W}(R,k)^2\,.
\end{equation}
In addition the skewness, $\skewR$, is calculated by employing the following expression
\begin{equation}\label{eq:skewness}
\skewR = \frac{\langle \Delta_R(\bx)^3 \rangle}{\langle \Delta_R(\bx)^2\rangle^{3/2}}\,.
\end{equation}
We shall, unless otherwise indicated, use $R=\sqrt{2}/k_{\rm peak}$, where $k_{\rm peak}$ represents the wavenumber approximately half an efold from the smallest scale on which the Gaussian amplitude is boosted (i.e. corresponding to $20$ grid points in Fourier space).

\section{Constraints on the small scale power spectrum}\label{sec:constraints}
Bounds on the abundance of PBHs, $\beta$, can be used to constrain the curvature perturbation power spectrum. Previous constraints have been obtained using an analytic method \cite{Shandera:2012ke,Byrnes:2012yx,Young:2013oia,Young:2014oea}, and it has been shown that the constraints can depend strongly on non-Gaussianity. It is normally assumed that PBHs form with approximately the horizon mass, although it is well known that the mass of the PBH that forms depends on the amplitude of the overdensity - and the mass has been found to follow a scaling law. The effect of this was recently considered \cite{Kuhnel:2015vtw} and leads to a shift and broadening of the PBH masses, and an overall decrease of the mass contained in primordial black holes. However, the PBHs formed still have approximately the horizon mass (the peak in the mass formed is typically half the horizon mass), and has a very small effect on the derived constraints - and so the effect is neglected here.

The effects of local-type non-Gaussianity have previously been studied, and it was found that the constraints on the power spectrum, $\mathcal{P}_\zeta$, can vary by up to an order of magnitude when $\fNLloc$ changes from $-0.5$ to $0.5$. Initially, a power spectrum which peaks over a small range of scales was considered \cite{Byrnes:2012yx,Young:2013oia}. Because $\fNLloc$ has a strong effect on the tails of the distribution function where PBHs form, small changes in $\fNLloc$ have a very large effect on the abundance of PBHs. Positive $\fNLloc$ increases the amount of PBHs which form such that the constraints become gradually tighter as $\fNLloc$ increases. For negative $\fNLloc$ the constraints loosen, but become weaker very quickly as $\fNLloc$ decreases, with no PBHs formed unless the power spectrum becomes much larger.

Later, the case where the power spectrum spans a larger range of scales was considered, allowing for the effect of super-horizon modes \cite{Young:2014oea}. Super-horizon modes normally do not directly affect PBH formation as far as is known \cite{Nakama:2014fra}, but can have an indirect effect due to modal coupling to horizon scale modes. Overall, the effect of modal coupling increases PBH formation - tightening constraints on the power spectrum. Notably, for negative $\fNLloc$, whilst constraints still weaken for small negative values, they become stronger as $\fNLloc$ becomes larger. A full discussion can be seen in \cite{Young:2014oea}.

The method detailed in section \ref{sec:sim_proc} is used to calculate the abundance of PBHs, $\beta$, as a function of the power spectrum and bispectrum - and this can be used to place an upper limit on the power spectrum for a given upper limit on $\beta$. Due to the amount of resources required to generate large maps, we restrict ourselves to a relatively weak constraint, $\beta<10^{-4}$. Whilst this constraint is weaker than any of the existing constraints on PBH abundance, it allows for an easier investigation of the effects of non-Gaussianity. It has been shown that the effect of non-Gaussianity upon the power spectrum constraint is relatively large compared to the effect of the constraint on $\beta$ \cite{Young:2014oea}. In any case, we expect the qualitative lessons drawn from our results to hold for any smaller value of $\beta$, although a simulation with a larger grid would have to be made to calculate the precise constraints.



\begin{figure}
 \begin{minipage}{1\textwidth}
\centering
\begin{subfigure}{0.5\textwidth}
	\centering
	\includegraphics[width=\linewidth]{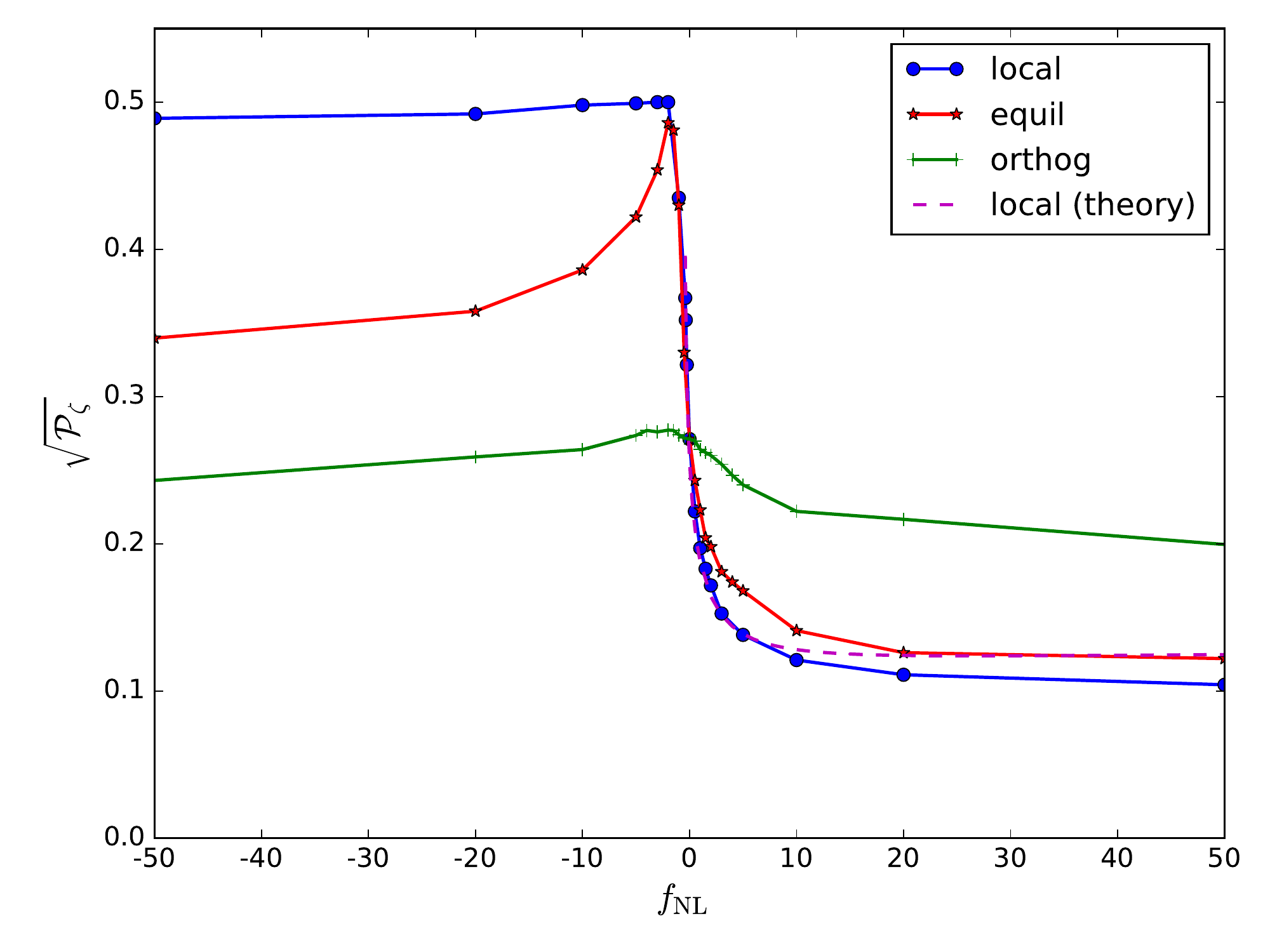}
\end{subfigure}%
\begin{subfigure}{0.5\textwidth}
	\centering
	\includegraphics[width=\linewidth]{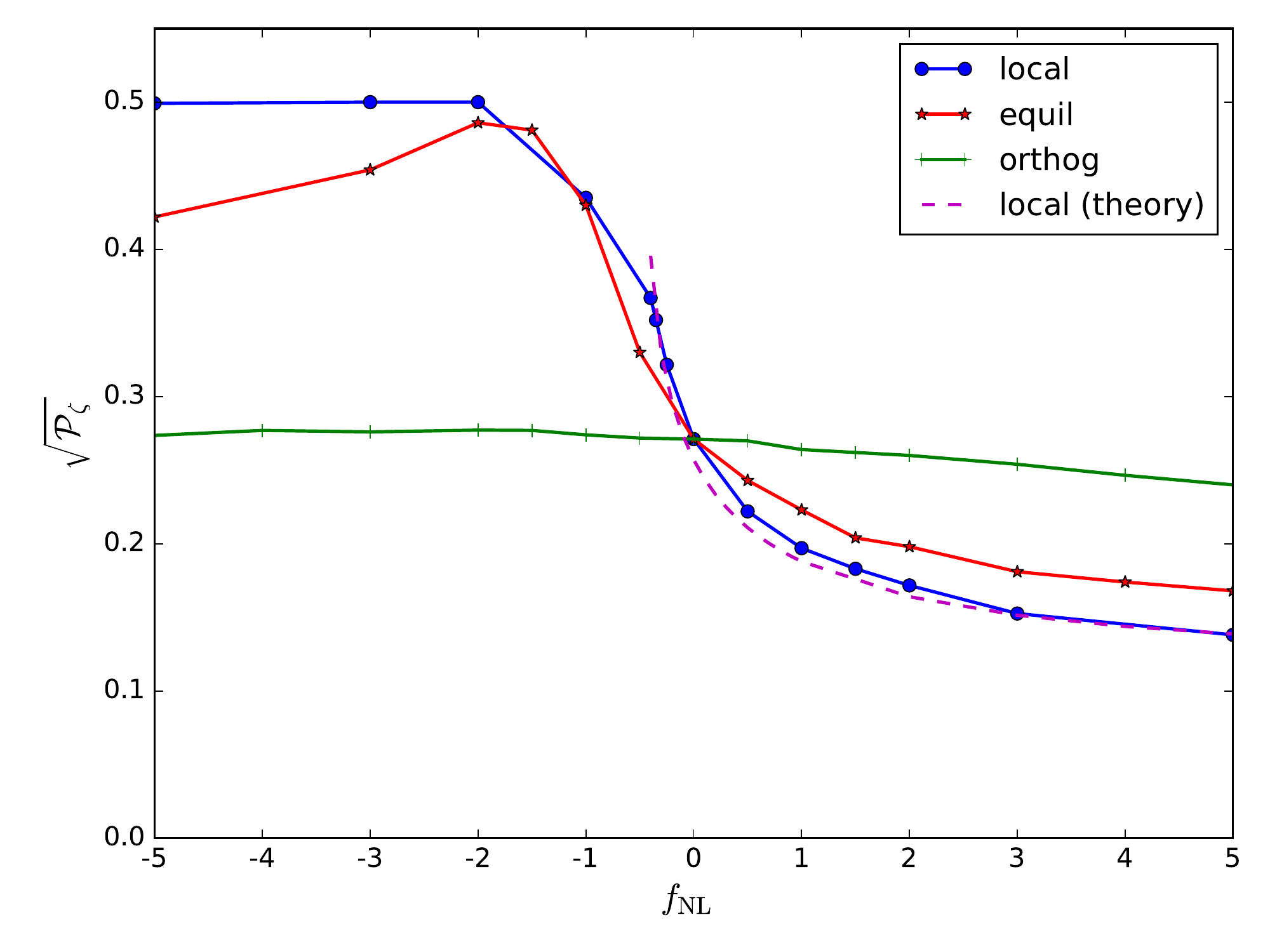}
\end{subfigure}
\caption{The constraints on the power spectrum as a function of $\fNL$ for the different bispectrum shapes is plotted. The plots show the upper limit on the power spectrum peak, spanning 1 e-fold, for a constraint on the abundance of PBHs $\beta<10^{-4}$. The right plot simply shows the central region of the left plot. Constraints become quickly tighter for positive $\fNL$ in the local and equilateral configurations, and weaker for negative $\fNL$. For the orthogonal configuration however, constraints are only weakly dependent on the value of $\fNLorth$. The dotted line represents the theoretical prediction for the constraint on the power spectrum for the local model originally derived in \cite{Byrnes:2012yx}. There is strong agreement for small values of $\fNLloc$, but the results disagree for larger values - although the same qualitative behaviour is seen.}
\label{1_efold_plots}  
 \end{minipage}
 \begin{minipage}{1\textwidth}
\centering
\begin{subfigure}{0.5\textwidth}
	\centering
	\includegraphics[width=\linewidth]{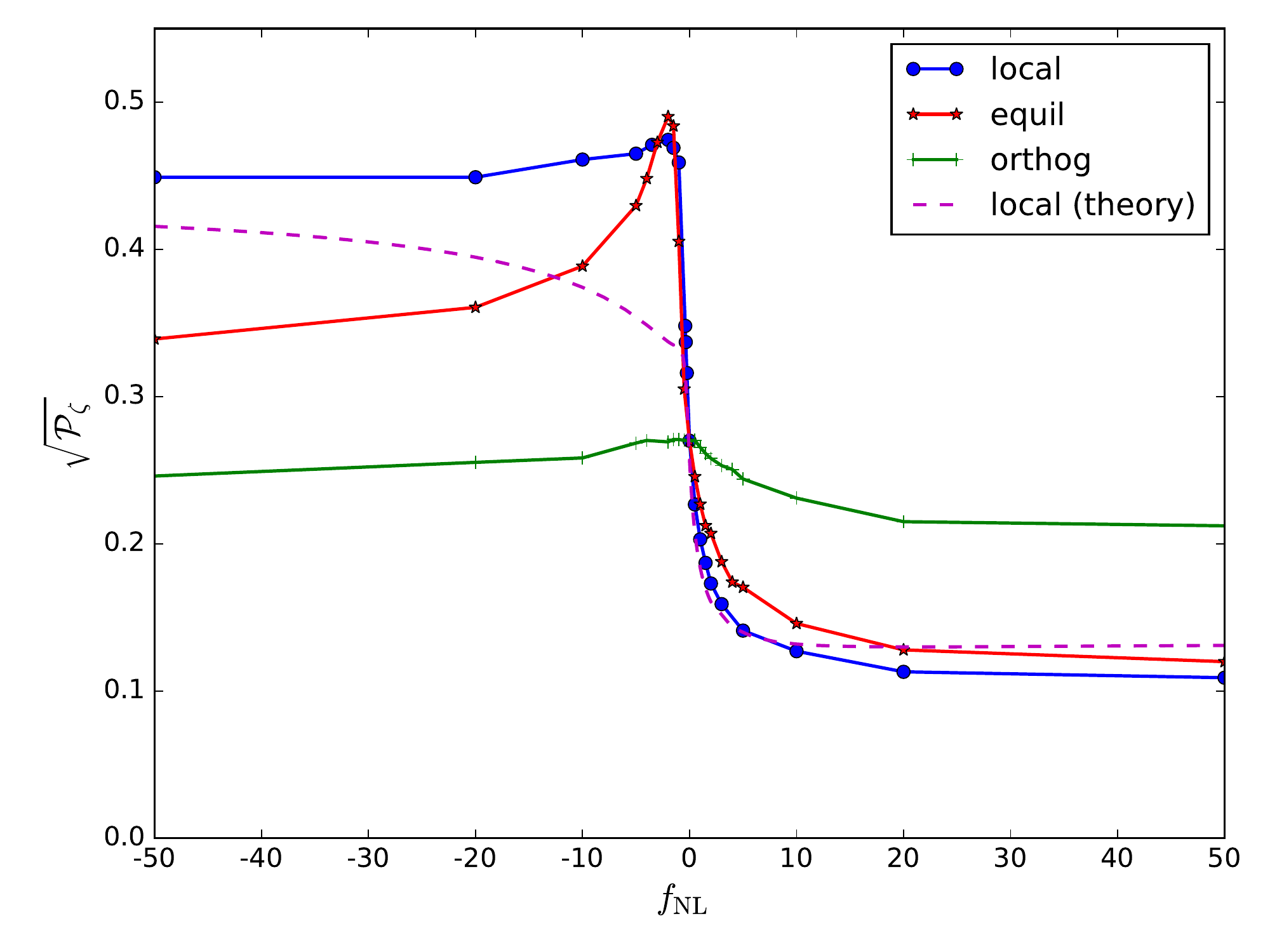}
\end{subfigure}%
\begin{subfigure}{0.5\textwidth}
	\centering
	\includegraphics[width=\linewidth]{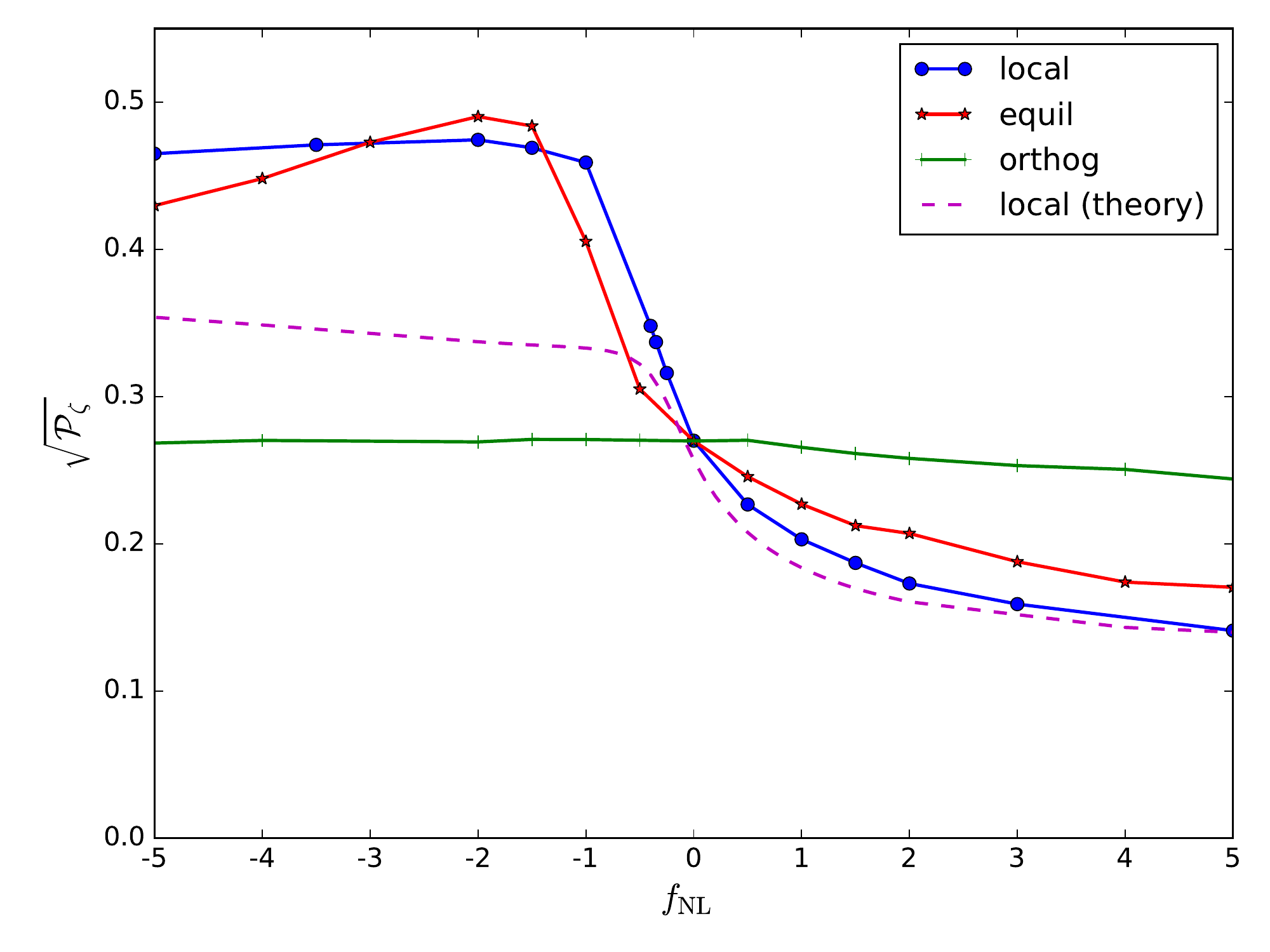}
\end{subfigure}
\caption{The constraints on the power spectrum as a function of $\fNL$ for the different bispectrum shapes is plotted. The plots show the upper limit on the power spectrum peak, spanning 2.5 e-folds, for a constraint on the abundance of PBHs $\beta < 10^{-4}$. The constraints display the same behaviour as seen in figure \ref{1_efold_plots}, with the exception that the constraints in the local model are slightly tighter due to stronger modal coupling now that different scale modes are being considered - especially for negative values of $\fNLloc$. As expected, the theoretical line for the local model does not match well for negative values - this is because the peak-background split has been used which assumes a large separation in scales between the ``peak" modes and the ``background" modes, with intermediate modes neglected.}
\label{25_efold_plots} \end{minipage}
\end{figure}

Figure \ref{1_efold_plots} shows the constraints on the peak value of the power spectrum, spanning 1 e-fold, obtained for different values of $\fNL$ for the local, equilateral and orthogonal bispectrum shapes, as well as the theoretical predictions for the local-type (as calculated in \cite{Young:2014oea}, with no super-horizon modes present).  The lines show the maximum allowed amplitude of the power spectrum given a constraint on the abundance of PBHs, $\beta<10^{-4}$. There is good agreement in the local model with the theoretical prediction for small values of $\fNLloc$, but mild disagreement for larger values - the theoretical model slightly overestimates the constraints for large positive $\fNLloc$. This is due to the fact that the calculation of the power spectrum assumes a dominant Gaussian component (there is much stronger agreement for the Gaussian component of the power spectrum). The bounds typically become stronger for positive values of $\fNL$ but significantly weaker for negative values. As $\fNL$ becomes large and negative, constraints quickly reach a maximum value before becoming slightly tighter - due to the effect of modal coupling \cite{Young:2014oea}. This is not seen for the theoretical prediction which does not account for the modal coupling - meaning the predicted constraints are much weaker. 

The exception is the orthogonal shape, with results showing that constraints on the power spectrum are relatively insensitive to orthogonal-type non-Gaussianity. This is due to the $\fNLorth$ having only a small effect on the skewness of the distribution, which will be discussed in more detail later in the paper. 

Another important note is that the theoretical calculation predicts the constraints on the power spectrum rapidly become weaker, and greater than unity, for negative $\fNL$ - and whilst the rapid weakening of constraints is still seen in the numerically generated constraints, they quickly reach some maximum value. In the case of equilateral- and orthogonal-type non-Gaussianity, the constraints then become stronger as more negative values of $\fNL$ are considered. This is believed to be due to the  strong signal in the bispectrum shapes when 3 modes of the same scale are considered - and so the effect of modal coupling tightens constraints, as discussed in more detail in \cite{Young:2014oea}. By contrast, the local-type peaks in the squeezed limit - when the modes considered are of significantly different scales - and so the effect of modal coupling is less important.

Figure \ref{25_efold_plots} shows the power constraints obtained for a peak in the power spectrum spanning 2.5 e-folds. Due to the computing resources required for a larger peak, we do not consider broader peaks than this in the power spectrum. The plot for the theoretical calculation for the local model now includes the effect of modal coupling to super-horizon modes with a large power spectrum spanning $\frac{1}{2}$ an e-fold \cite{Young:2014oea}. The constraints obtained are similar to the case where a narrower peak in the power spectrum is considered - positive $\fNL$ increases PBH abundance and tightens constraints, whilst negative $\fNL$ has the opposite effect. 

By eye, figures \ref{1_efold_plots} and \ref{25_efold_plots} look very similar - although constraints are tighter for non-zero $\fNL$, the most significant difference can be seen in the local model. Changing the smoothing scale, or equivalently adding in super-horizon modes, does not signficantly affect the tail of the probability distribution function of the density contrast for equilateral- or orthogonal-type non-Gaussianity, as the horizon-scale curvature perturbation modes are largely independent of super-horizon modes. The constraints obtained for these models are therefore approximately equal between figures \ref{1_efold_plots} and \ref{25_efold_plots}. By contrast, local-type non-Gaussianity displays a strong modal coupling to super-horizon scales, and we see a larger difference between the constraints when the smoothing scale is changed. The difference is, however, still small because the range of scales being considered is still relatively small. A larger difference would be seen if a much broader peak in the power spectrum is considered, as is shown analytically in \cite{Young:2014oea} for the local model, where a significant change in the tail of the distribution function is seen due to the fact that the local shape peaks in the squeezed limit (where the modes are of significantly different scales). The peak in the power spectrum is now broad enough that small scale modes and large scale modes are considered - allowing the effect of significant modal coupling. It has previously been noted that modal coupling typically increases PBH production, and tightens constraints \cite{Young:2014oea}. Thus, when a broader peak in the power spectrum is considered, constraints become tighter for the local shape - but remain largely unchanged for the equilateral and orthogonal shapes.

\subsection{Skewness}

We will now consider the skewness of the different bispectrum shapes, and show that when the non-Gaussianity and skewness parameters, $\fNL$ and $\skewR$, respectively, are small that the skewness alone can be considered to produce constraints on the power spectrum. However, as $\fNL$ and the skewness become large, the effects of the different bispectrum shapes must be considered. The skewness, given by equation~\eqref{eq:skewness}, may be computed for a given bispectrum using, 
\begin{align}
\langle \Delta_R(\bx)^2 \rangle&=\int d\ln k {W}(R,k)^2 \frac{k^3 P_{\zeta}(k)}{2\pi^2} \,,\nonumber\\
\langle \Delta_R(\bx)^3 \rangle&=\frac{2}{(2\pi)^4}\int\limits_0^\infty d\mathrm{ln}(k) k^3 {W}(R,k) \int\limits_0^\infty d\mathrm{ln}(q) q^3 {W}(R,q) \int\limits_{-1}^1 d\mu\, {W}(R,k_\mu){B_\zeta}(k,q,k_\mu)\,,
\end{align}
where $\mu=\mathrm{cos}(\theta)$, with $\theta$ representing the angle between ${\bf k}$ and ${\bf q}$. Calculating the skewness for the three bispectrum shapes being considered using this formula gives:
\begin{equation}
\skewR^{\rm local}=2.6\fNLloc\sqrt{\mathcal{P}_\zeta}\, ,
\end{equation}
\begin{equation}
\skewR^{\rm equil}=1.1\fNLeq\sqrt{\mathcal{P}_\zeta}\, ,
\end{equation}
\begin{equation}
\skewR^{\rm orthog}=0.07\fNLorth\sqrt{\mathcal{P}_\zeta}\, .
\end{equation}
We note that the numbers obtained here are slightly different than the values obtained by Shandera et al for the local and equilateral model \cite{Shandera:2012ke}, due to the choice of window functions, transfer functions and the form of the power spectrum. To our knowledge, this is the first the skewness has been calculated for the orthogonal model.

We see that the skewness is relatively large for the local and equilateral shapes but small for the orthogonal shape - which is why the constraints are less dependent on $\fNLorth$ than on $\fNLloc$ and $\fNLeq$. Note that the above analytic formulae are only correct whilst the skewness is small. 

\begin{figure}[t]
\centering
\includegraphics[width=0.7\linewidth]{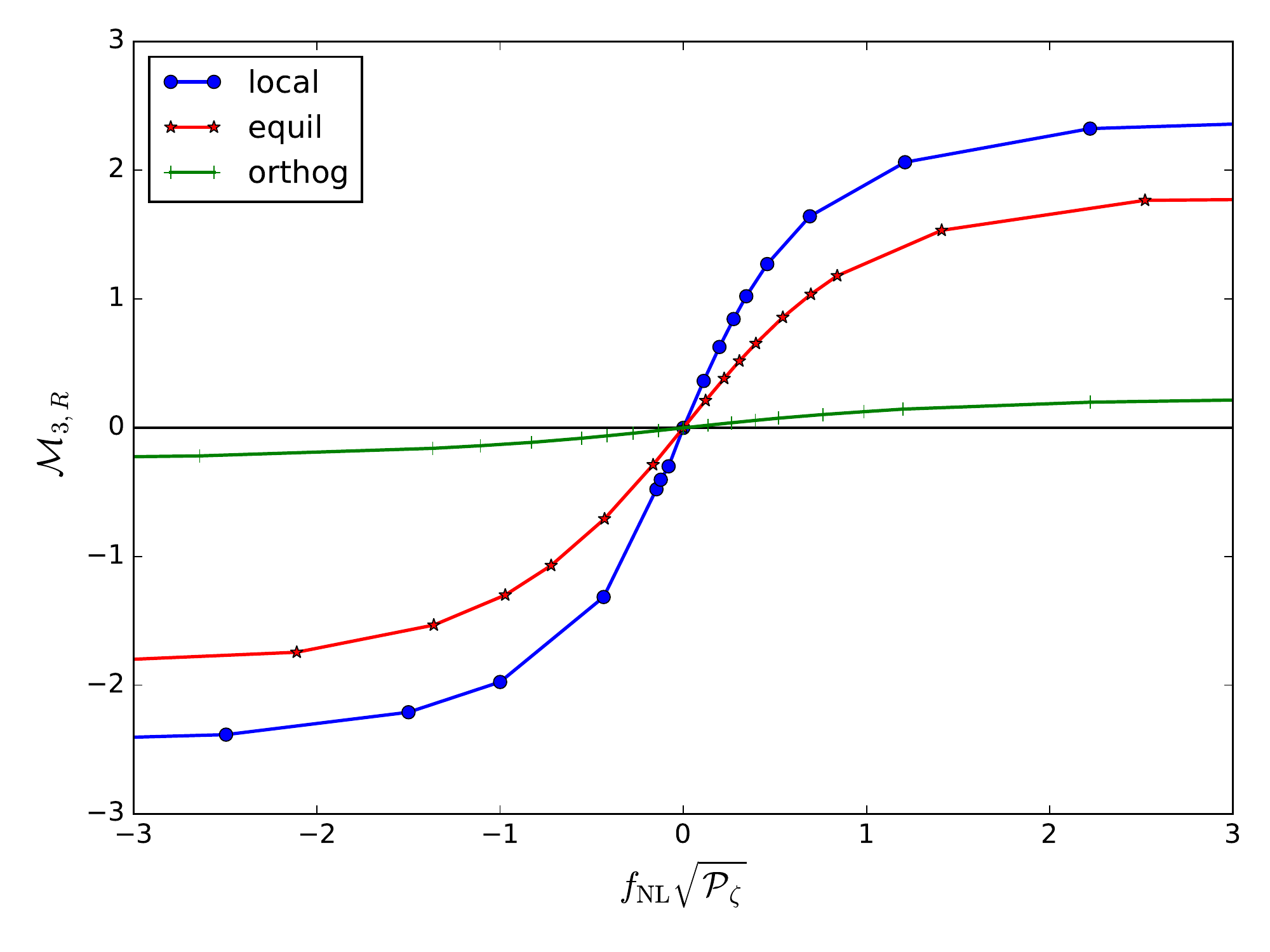}
\caption{The skewness, $\skewR$, is plotted against $\fNL\sqrt{\mathcal{P}_\zeta}$ for an abundance of PBHs $\beta=10^{-4}$. It can be seen that the skewness does not depend linearly on $\fNL\sqrt{\mathcal{P}_\zeta}$, unlike that predicted by equation \eqref{eq:skewness}. However, in the central region where the skewness is small, $|\skewR |<\mathcal{O}(0.1)$, the relation is approximately linear, and the skewness may be used to parameterise the abundance of PBHs and constraints on the power spectrum.}
\label{skewness_fixed_beta}
\end{figure}

Figure \ref{skewness_fixed_beta} shows how the skewness varies as a function of $\fNL\sqrt{\mathcal{P}_\zeta}$. The plot is generated from the simulated density maps for a fixed abundance of PBHs, $\beta=10^{-4}$. Confirming the above calculation, the skewness is seen to be the largest for local non-Gaussianity, and smallest for orthogonal non-Gaussianity. The relation is also strongly non-linear as the skewness becomes large - which indicates the region where skewness can no longer be used to parameterise the abundance of PBHs. The skewness saturates relatively quickly as $\fNL$ increases - representing the fact that the distribution has become dominated by the non-Gaussian components. The fact that the skewness reaches some constant value as $\fNL$ becomes larger also corresponds to the fact that the constraints asymptote to a constant level as $\fNL$ becomes larger.

\begin{figure}[t]
\centering
\begin{subfigure}{0.5\textwidth}
	\centering
	\includegraphics[width=\linewidth]{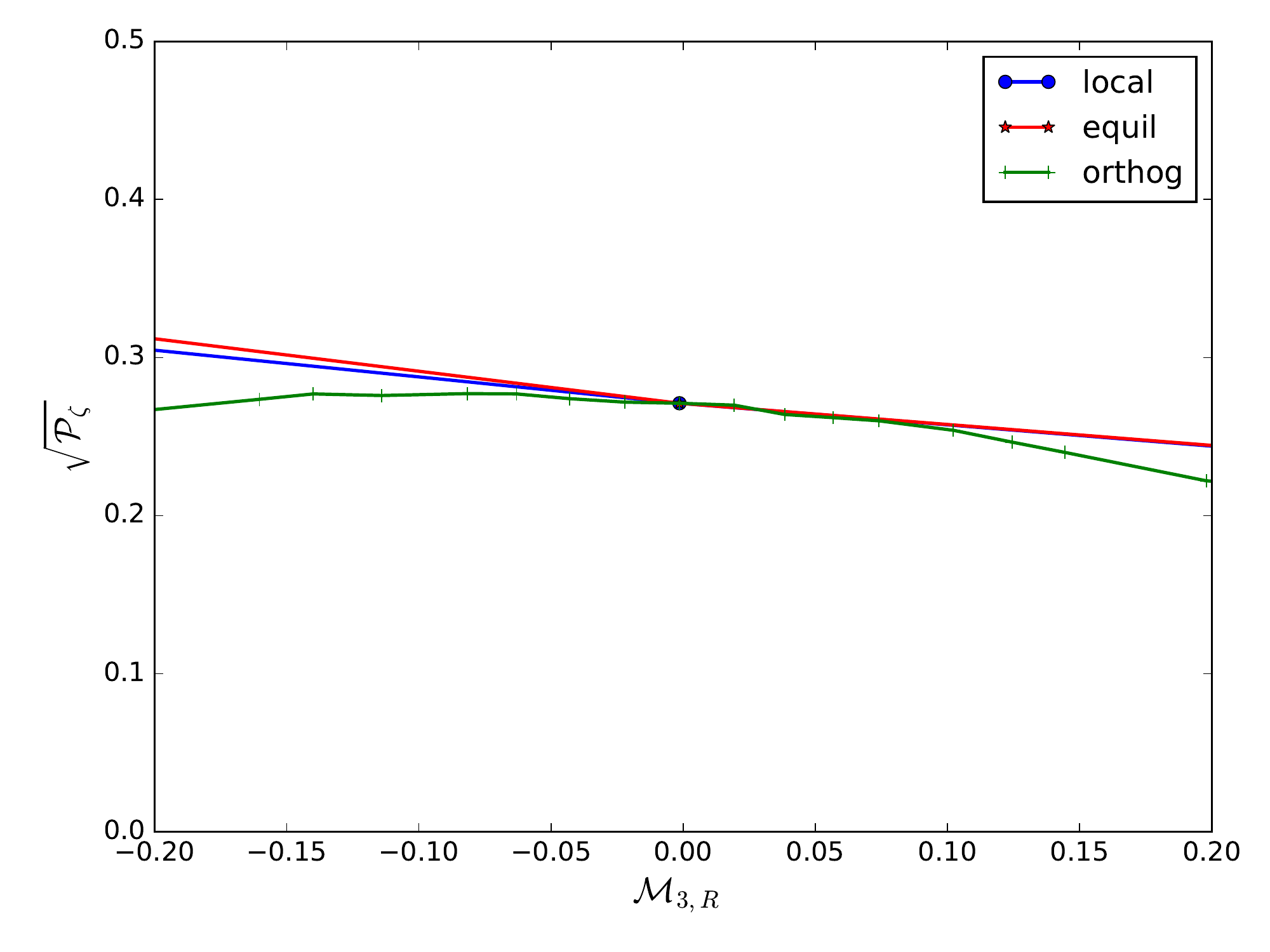}
\end{subfigure}%
\begin{subfigure}{0.5\textwidth}
	\centering
	\includegraphics[width=\linewidth]{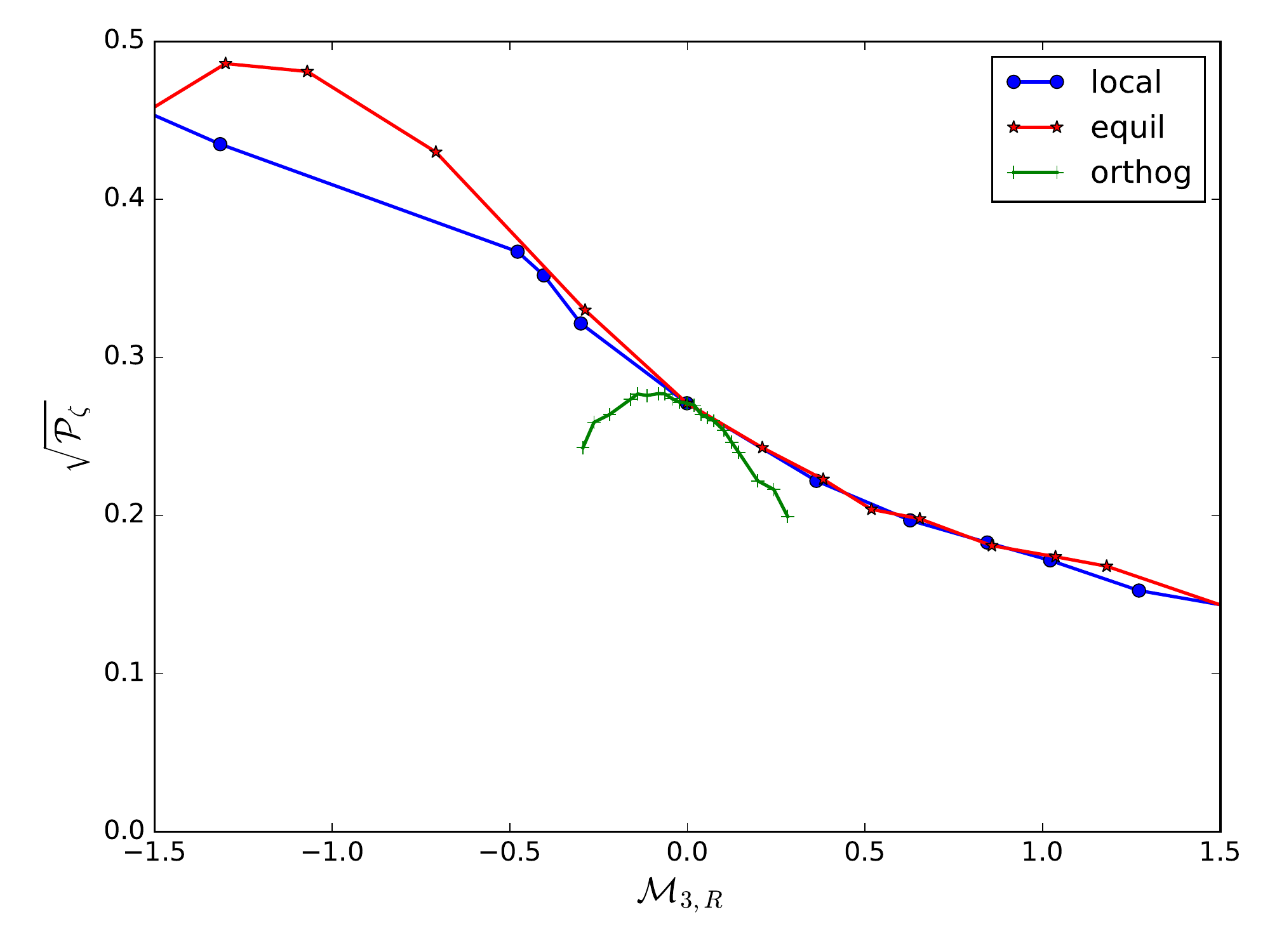}
\end{subfigure}
\caption{The constraint on the power spectrum corresponding to a constraint on the abundance of PBHs, $\beta<10^{-4}$, as a function of the skewness is plotted. As can be seen in the left plot, for the three bispectrum shapes considered, the constraints on the power spectrum show good agreement whilst the skewness is small, $\skewR<\mathcal{O}(0.1)$. The skewness of the distribution is therefore the most important consideration, rather than the shape of the bispectrum. The right plot shows the behaviour as the skewness becomes large - the shape of the bispectrum has a large impact on the derived constraints and this must therefore be taken into account.}
\label{skew_vs_power}
\end{figure}

Figure \ref{skew_vs_power} plots the upper bound on the power spectrum corresponding to a constraint on the abundance of PBHs, $\beta<10^{-4}$, as a function of the skewness. Whilst the skewness is small, $\skewR<\mathcal{O}(0.1)$, the skewness of the distribution is the most important consideration, rather than the shape of the bispectrum. This can be seen in the left plot of figure \ref{skew_vs_power}. However, as the non-Gaussianity, $\fNL$, and the skewness, $\skewR$, become larger, this is no longer the case - and a large discrepancy between the different bispectrum configurations can be seen the right plot.

\section{Summary}\label{sec:conclusions}
The lack of observation of PBHs allows tight constraints to be placed on the mass fraction of the universe collapsing into PBHs at the time of formation, $\beta$. This, in turn, allows unique bounds to be placed on the small scale primordial curvature perturbation power spectrum, $\mathcal{P}_\zeta$, at scales which are otherwise unobservable - although these bounds are orders of magnitude weaker than constraints from sources such as the CMB. Non-Gaussian density maps were generated and used to predict the abundance of PBHs for different shapes of bispectrum, in the local, equilateral and orthogonal configurations. These predictions were then used to place constraints on $\mathcal{P}_\zeta$ as a function of the amplitude and shape of the bispectrum.

As an improvement on previous work, this method allows the consideration of bispectra of arbitrary shape and amplitude. We confirmed the previous findings using analytic methods of the effects of local-type non-Gaussianity \cite{Byrnes:2012yx,Young:2013oia,Young:2014oea} - non-Gaussianity can have a strong effect on constraints on the power spectrum, typically becoming stronger (weaker) for positive (negative) values of $\fNL$. The effect of the skewness was also considered, confirming results seen in \cite{Shandera:2012ke} ( note that \cite{Clark:2015tha} also recently used a similar technique to calculate constraints arising from ultra-compact mini-haloes) - but demonstrate that using the skewness to parameterise the abundance of PBHs is only valid for small amounts of non-Gaussianity. As seen in figure \ref{skew_vs_power}, for small amounts of skew the shape of the bispectrum has little effect on the constraints - and the skewness of the distribution can be considered the most important factor (but note that the constant of proportionality relating $\fNL$ to the skewness does strongly depend on the non-Gaussian template). However, for large non-Gaussianity, constraints on the power spectrum become strongly dependent on the shape of the bispectrum. 

For the local and equilateral shapes the constraints become tighter for positive $\fNL$ but dramatically weaker for small negative $\fNL$. For orthogonal-type non-Gaussianity, the effects are qualitatively similar, but much less dramatic - due to the relatively small skewness generated by this bispectral shape. Previous findings that the effect of modal coupling and positive skew is to increase PBH formation, whilst negative skew decreases PBH formation, are also confirmed.

\section*{Acknowledgements} SY is supported by an STFC studentship, and CB is supported by a Royal Society University
Research Fellowship. DR acknowledges support from the European Research Council under the European Union's Seventh Framework Programme (FP/2007-2013) / ERC Grant Agreement No. [308082].

\bibliographystyle{JHEP}
\bibliography{bibfile}

\end{document}